# Influence of Ru composition deviation from stoichiometry on intrinsic spin-to-charge conversion in SrRuO$_3$

Shingo Kaneta-Takada,[1,*] Yuki K. Wakabayashi,[2] Hikari Shinya,[1,3,4,5,6] Yoshitaka Taniyasu,[2] Hideki Yamamoto,[2] Yoshiharu Krockenberger,[2] Masaaki Tanaka,[1,3,7] and Shinobu Ohya[1,3,7*]

[1]*Department of Electrical Engineering and Information Systems, The University of Tokyo, Bunkyo, Tokyo 113-8656, Japan*
[2]*NTT Basic Research Laboratories, NTT Corporation, Atsugi, Kanagawa 243-0198, Japan*
[3]*Center for Spintronics Research Network (CSRN), The University of Tokyo, Bunkyo, Tokyo 113-8656, Japan*
[4]*Center for Spintronics Research Network (CSRN), Graduate School of Engineering Science, Osaka University, Toyonaka, Osaka 560-8531, Japan*
[5]*Center for Science and Innovation in Spintronics (CSIS), Tohoku University, Sendai, Miyagi 980-8577, Japan*
[6]*Center for Spintronics Research Network (CSRN), Institute for Chemical Research, Kyoto University, Uji, Kyoto 611-0011, Japan*
[7]*Institute for Nano Quantum Information Electronics, The University of Tokyo, 4-6-1 Meguro-ku, Tokyo 153-8505, Japan*
[*]Email: skaneta@cryst.t.u-tokyo.ac.jp, ohya@cryst.t.u-tokyo.ac.jp

**Abstract:** Interconversion between charge and spin currents is a key phenomenon in realizing next-generation spintronic devices. Highly efficient spin-charge interconversion is expected to occur at band crossing points in materials with large spin-orbit interactions due to enhanced spin Berry curvature. On the other hand, if defects and/or impurities are present, they affect the electronic band structure, which in turn reduces the spin Berry curvature. Although defects and impurities are generally numerous in materials, their influence on the spin Berry curvature and, consequently, spin-charge interconversion has often been overlooked. In this paper, we perform spin-pumping experiments for stoichiometric SrRuO$_3$ and non-stoichiometric SrRu$_{0.7}$O$_3$ films at 300 K, where the films are in paramagnetic states, to examine how Ru composition deviation from the stoichiometric condition influences the spin-to-charge conversion, showing that SrRuO$_3$ has a larger spin Hall angle than SrRu$_{0.7}$O$_3$. We derive the band structures of paramagnetic SrRuO$_3$ and SrRu$_{0.75}$O$_3$ using first-principles calculations, indicating that the spin Hall conductivity originating from the spin Berry curvature decreases when the Ru deficiency is incorporated, which agrees with the experimental results. Our results suggest that point-defect- and impurity control is essential to fully exploit the intrinsic spin Berry curvature and large spin-charge interconversion function of materials. These insights help us with material designs for efficient spin-charge interconversions.

# I. INTRODUCTION

Interconversion phenomena between charge and spin currents [1,2] hold great promise for the evolution of next-generation efficient spintronics devices, such as spin-orbit torque magnetoresistive random access memory [3,4], magnetoelectric spin-orbit logic devices [5], and spin Seebeck power generators [6]. Spin-to-charge conversions are classified into two phenomena: the inverse spin Hall effect in bulk [2] and the inverse Rashba-Edelstein effect at material surfaces or interfaces [7,8]. For the inverse spin Hall effect, three origins have been proposed depending on the conductivities [9]: extrinsic side-jump scattering [10,11], extrinsic skew scattering [12,13], and intrinsic scattering due to spin Berry curvature [14]. In the intrinsic case, when band crossing points exist in an electronic band structure, highly efficient spin-charge interconversions occur due to their large spin Berry curvatures, as demonstrated for heavy-metal Pt and topological materials [15]. In those cases, however, if defects and impurities are present, they affect the band structure and thus the spin Berry curvature. Generally, materials have numerous defects and impurities. Thus far, few studies have focused on how they affect the spin Berry curvature and the spin-charge interconversion in materials with band crossings.

The itinerant $4d$ perovskite oxide $SrRuO_3$ has recently attracted much attention in the spintronics research field, primarily owing to its distinctive features, including bulk three-dimensional Weyl fermions [16,17] and surface Fermi arcs [18,19]. The band crossing points in $SrRuO_3$ show large spin Berry curvatures [20,21], resulting in highly efficient spin-charge interconversions [22]. Importantly, previous studies have clarified that Ru stoichiometry is crucial for clear observation of the Weyl states using quantum transport measurements; its deviation alters the magnetic and (quantum) transport properties, hindering the Weyl fermion transport [23,24,25].

Here, we conduct spin-pumping experiments to investigate the influence of the Ru stoichiometry change on the spin-to-charge conversion in $SrRuO_3$ and $SrRu_{0.7}O_3$ films at 300 K, where the films are in paramagnetic states. $SrRuO_3$ shows a higher spin-to-charge conversion-induced electromotive force than $SrRu_{0.7}O_3$. From this result, we find that $SrRuO_3$ has a larger spin Hall angle than $SrRu_{0.7}O_3$. Furthermore, we calculate the electronic band structures of $SrRuO_3$ and $SrRu_{0.7}O_3$ using first-principles calculations to theoretically estimate the spin Hall conductivity originating from the spin Berry curvature. We find that stoichiometric $SrRuO_3$ has a large spin Hall conductivity with a large spin Berry curvature at a $k$ point where a band crossing appears near $E_F$, and it decreases with the composition deviation of Ru from the stoichiometric condition, which agrees with our experimental results. Our findings suggest that the deviation from the stoichiometry in materials alters the spin Berry curvature and the spin-charge interconversion efficiencies. These insights will be useful for material designs to realize efficient spin-charge interconversions.

## II. SAMPLE PREPARATION AND EXPERIMENTAL SETUP

For the experiments, we prepared all-epitaxial oxide heterostructures composed of $(La_{0.67},Sr_{0.33})MnO_3$ (LSMO) (12 nm, ferromagnetic spin injection layer)/$SrRu_{1-v}O_3$ ($v$ = 0 or 0.3) (10 nm, paramagnetic spin-to-charge conversion layer) on $TiO_2$-terminated $SrTiO_3$ (001) substrates using molecular beam epitaxy (MBE) (Fig. 1). The $SrRu_{1-v}O_3$ layer was grown with a machine-learning-assisted MBE setup [26,27] equipped with multiple e-beam evaporators for Sr and Ru [28,29,30]. Oxidation during the growth was carried out with a mixture of ozone ($O_3$) and oxygen ($O_2$) gas (~15% $O_3$ + 85% $O_2$), which was introduced through a nozzle pointed at the substrate at a flow rate of ~2 sccm. The growth temperature was fixed at 772 °C for the $SrRu_{1-v}O_3$ layer. The Ru flux rates were set at 0.365 and 0.190 Å/s for the growth of the stoichiometric $SrRuO_3$ and Ru-deficient $SrRu_{0.7}O_3$ films, respectively, while the Sr flux was fixed at 0.98 Å/s [23,24,25]. When we grew the stoichiometric $SrRuO_3$ layer, we used an excessive Ru flux because excessive Ru atoms evaporate from the growth surface forming volatile species such as $RuO_3$ and $RuO_4$ under an oxidizing atmosphere, which enables us to make stoichiometric films. Using energy-dispersive X-ray spectroscopy and X-ray photoemission spectroscopy measurements, we estimated the composition ratios of Ru/Sr for the $SrRuO_3$ and $SrRu_{0.7}O_3$ layers to be 1.03 and 0.71, respectively [24].

After the growth, the samples were exposed to air and transferred to another MBE chamber for the growth of LSMO. Before the growth of LSMO layers, the surface of the $SrRu_{1-v}O_3$ layer was cleaned by heating the samples at 720 °C and introducing a mixture of $O_3$ and $O_2$ (~20% $O_3$ + 80% $O_2$) up to a background pressure of $2 \times 10^{-4}$ Pa. The reflection high-energy electron diffraction (RHEED) pattern of the $SrRu_{1-v}O_3$ layer was recovered through this process as shown in Figs. 2(a) and 2(b), which were taken just before the growth of LSMO with an electron beam azimuth along the [100] direction of the $SrTiO_3$ substrate. We used shuttered growth for growing the LSMO layer with fluxes of La, Sr, and Mn supplied by Knudsen cells [8,31,32,33]. We controlled the layer thickness monitoring the oscillation of *in-situ* RHEED. The LSMO layer was grown at 720 °C with a background pressure of $2 \times 10^{-4}$ Pa due to a mixture of $O_3$ and $O_2$ (~20% $O_3$ + 80% $O_2$). Figures 2(a) and 2(b) also show the RHEED patterns with sharp streaky patterns obtained for the LSMO layer in the LSMO/$SrRuO_3$ and LSMO/$SrRu_{0.7}O_3$ samples, respectively, indicating the epitaxial growth of LSMO/SrRu$_{1-v}$O$_3$ on SrTiO$_3$ substrates despite the exposure to air for transferring the samples.

In the $\theta$–$2\theta$-scan X-ray diffraction (XRD) pattern around the SrTiO$_3$ (002) reflection peak, we see peaks of LSMO, SrRuO$_3$, and SrRu$_{0.7}$O$_3$ with clear Laue fringes, indicating the formation of sharp interfaces/surfaces of the films [Fig. 2(c)]. Based on the (002) peak positions, the out-of-plane lattice constant of SrRu$_{0.7}$O$_3$ is slightly larger than that of SrRuO$_3$. This result is consistent with the previous reports claiming that Ru vacancies increase the out-of-plane lattice constant and unit cell volume of SrRuO$_3$ films on SrTiO$_3$ substrates [24,34]. Figure 2(d) shows the reciprocal space mapping (RSM) around SrTiO$_3$ ($\bar{1}$03). The horizontal peak positions of LSMO and SrRuO$_3$ are identical to that of the SrTiO$_3$ substrate, confirming that the film is grown coherently. We note that the horizontal peak positions of SrRu$_{0.7}$O$_3$ and LSMO are also identical to those of the SrTiO$_3$ substrate (not shown).

For the transport measurements using a physical property measurement system (PPMS) of Quantum Design, we cut the samples into a rectangular shape. We adopted a standard four-probe method with indium electrodes. For the magnetization measurements, we used a magnetic property measurement system (MPMS-3) of Quantum Design. Spin pumping measurements were conducted using a transverse electric TE$_{011}$ cavity in an electron spin resonance system with a microwave frequency of 9.1 GHz and with various microwave powers (*MPs*) ranging from 10 to 70 mW. The width *w* and length *l* of the samples used for spin-pumping experiments were 0.8 mm and 2.2 mm for LSMO/SrRuO$_3$ and 1.2 mm and 1.5 mm for LSMO/SrRu$_{0.7}$O$_3$, respectively.

Electronic band structure calculations were performed using the QUANTUM ESPRESSO package [35,36,37]. We used a plane-wave basis and a pseudopotential from the PSlibrary [38]. We used a fully relativistic pseudopotential with the generalized gradient approximation based on the projector-augmented plane wave method with the Perdew-Burke-Ernzerhof functional [39,40]. The plane-wave and charge-density cutoff energies were set to 52 and 368 Ry, respectively. The Brillouin zone was sampled with 4×4×4 meshes, and the convergence criterion for the electron density was set at $1 \times 10^{-8}$ eV in the self-consistent calculations. The experimental crystal structure of orthorhombic SrRuO$_3$ was adopted for all calculations (the *Pbnm* space group, *a* = 5.5670 Å, *b* = 5.5304 Å, *c* = 7.8446 Å) [41]. To calculate the electronic band structure of SrRu$_{0.75}$O$_3$, we performed lattice relaxation in a single unit cell of orthorhombic SrRuO$_3$ with a Ru atom removed. For the theoretical calculation of the spin Hall conductivity, the electronic band structure was interpolated with the maximally localized Wannier functions by projecting the bands near the Fermi level onto the Ru *d* and O *p* atomic orbitals, as implemented in the WANNIER90 package [42,43,44,45]. The theoretical spin Hall conductivity and spin Berri curvature were calculated by using the WannierBerri package [46,47,48].

### III. RESULTS AND DISCUSSION

We measured the temperature $T$ dependence of the sheet resistance $R_{sheet}$ for the LSMO/SrRuO$_3$ and LSMO/SrRu$_{0.7}$O$_3$ films (Fig. 3). The LSMO/SrRu$_{0.7}$O$_3$ sample has a higher $R_{sheet}$ than LSMO/SrRuO$_3$ over the entire temperature range, indicating that Ru defects act as electron scatterers. The kinks highlighted by arrows in Fig. 3(a) correspond to the Curie temperature $T_C$ of SrRu$_{1-v}$O$_3$, below which spin-dependent scattering is suppressed due to the ferromagnetic transition, lowering the resistivity. From the maximum values of $dR/dT$ [Fig. 3(b)], the $T_C$ values are estimated to be 144 and 136 K for SrRuO$_3$ and SrRu$_{0.7}$O$_3$, respectively. The lower $T_C$ in SrRu$_{0.7}$O$_3$ indicates that Ru deficiency weakens the magnetic coupling in SrRuO$_3$.

The LSMO layer in our samples has $T_C$ above 300 K, as shown in the $T$ dependence of the magnetization $M$ [Fig. 4(a)]. We note that the $T_C$ of the SrRu$_{1-v}$O$_3$ layers is lower than the temperature range used in Fig. 4(a), and thus, $M$ shown here purely originates from the LSMO layer. We observe clear hysteresis loops of LSMO when the magnetic field $H$ is applied along the in-plane [110] direction of SrTiO$_3$, indicating that the easy magnetization axis of LMSO is along this direction [Fig. 4(b)].

To estimate the spin Hall angle for the LSMO/SrRuO$_3$ and LSMO/SrRu$_{0.7}$O$_3$ samples, we carried out spin-pumping measurements using ferromagnetic resonance (FMR). Here, the spin current is injected from ferromagnetic LSMO into paramagnetic SrRuO$_3$ or SrRu$_{0.7}$O$_3$, where the spin current is converted to a charge current (Fig. 1). For the measurements, a static magnetic field $H$ was applied along the [110] direction of the SrTiO$_3$ substrate in the film plane (*i.e.*, $\theta_H = 0°$ or 180°), which corresponds to the easy magnetization axis of LSMO. Here, $\theta_H$ is defined as the out-of-plane angle of $\mu_0H$ with respect to the in-plane [110] axis. The microwave magnetic field $h_{rf}$ was applied along the [1$\bar{1}$0] direction in the film plane. For the analysis, we subtracted the background by using a polynomial function up to the ninth order from the raw data of the $\mu_0H$ dependence of the electromotive force. As shown in the $\mu_0H$ dependence of $V$ for various *MP*s at 300 K with $\theta_H = 0$ and 180° [Figs. 5(a) and 5(b)], $V$ is enhanced at the FMR field $\mu_0H_{FMR}$ (=215 mT), and $V$ monotonically increases with increasing *MP*, which are typical results of spin-pumping experiments. The most noticeable point is that the magnitude of $V$ in the LSMO/SrRuO$_3$ sample is significantly larger than that in the LSMO/SrRu$_{0.7}$O$_3$ sample.

Before estimating the spin Hall angle, we extracted the inverse spin Hall signal from the data, as described below. We decompose the $V$ signals in Figs. 5(a) and 5(b) into a symmetric component $V_{sym}$, which includes the inverse spin Hall signal, and an asymmetric component $V_{asym}$ using the following equation:

$$V(H) = V_{sym} \frac{\Delta H^2}{(H - H_{FMR})^2 + \Delta H^2} + V_{asym} \frac{-2\Delta H(H - H_{FMR})}{(H - H_{FMR})^2 + \Delta H^2}, \quad (1)$$

where $\Delta H$ is the half-width at half-maximum of the FMR linewidth. As shown in Fig. 5(c), the obtained $V_{sym}$ shows a linear relation with the $MP$, a typical result of spin-pumping experiments. To extract the signal of the inverse spin Hall effect, whose sign should be reversed with a reversal in the $H$ direction, and to eliminate the contribution of the Seebeck effect, we derived $V_{sym,ave}$, which is defined as $(V_{sym,0°} - V_{sym,180°})/2$, where $V_{sym,0°}$ and $V_{sym,180°}$ are the $V_{sym}$ values for $\theta_H = 0°$ and 180°, respectively [49]. Next, we obtained the mixing conductance $g_{r,i}^{\uparrow\downarrow}$ ($i$ = LSMO/SrRuO$_3$ and LSMO/SrRu$_{0.7}$O$_3$) from the difference in the FMR peak-to-peak widths between the LSMO/SrRu$_{1-v}$O$_3$ sample ($\Delta H_{pp,i}$) and a reference LSMO sample reported in Ref. [50] ($\Delta H_{pp,LSMO}$) where spin current injection is negligible:

$$g_{r,i}^{\uparrow\downarrow} = \frac{4\pi M_S d_{LSMO}}{g\mu_B}(\alpha_i - \alpha_{LSMO}), \quad \text{where} \quad \alpha_i = \frac{\sqrt{3}\gamma \Delta H_{pp,i}}{2\omega}$$

$$\text{for } i = \text{LSMO/SrRuO}_3, \text{LSMO/SrRu}_{0.7}\text{O}_3, \text{ and LSMO.} \quad (2)$$

Here, $M_S$, $d_{LSMO}$, $g$, $\mu_B$, $\alpha_i$, $\gamma$, and $\omega$ are the saturation magnetization of the LSMO, the thickness of the LSMO layer, $g$-factor [51], Bohr magneton, damping constant, gyromagnetic ratio, and angular frequency, respectively. Then, spin current density $j_S^i$ is expressed by

$$j_S^i = \frac{g_r^{\uparrow\downarrow} \gamma^2 h_{rf}^2 \hbar [4\pi M_S \gamma + \sqrt{(4\pi M_S)^2 \gamma^2 + 4\omega^2}]}{8\pi \alpha_i^2 [(4\pi M_S)^2 \gamma^2 + 4\omega^2]}, \quad (3)$$

where $\hbar$ is the Dirac constant. The estimated $j_S^i$ was 7.45×10$^{-12}$ J/m$^2$ for the LSMO/SrRuO$_3$ sample and 1.71×10$^{-11}$ J/m$^2$ for the LSMO/SrRu$_{0.7}$O$_3$. Finally, we estimated the $\theta_{SHE}$ using

$$V_{sym,ave} = \frac{l\theta_{SHE}\lambda\tanh(d_{SRO}/2\lambda)}{\sigma_{LSMO}d_{LSMO} + \sigma_{SRO}d_{SRO}}\left(\frac{2e}{\hbar}\right)j_S^i, \quad (4)$$

where $\sigma_{SRO}$, $\sigma_{LSMO}$, $d_{SRO}$, and $d_{LSMO}$ are the conductivities and thicknesses of the SrRu$_{1-v}$O$_3$ and LSMO layers; $l$ is the length of the sample; $e$ is the elementary charge; $\lambda$ is the spin diffusion length of SrRu$_{1-v}$O$_3$. In this estimation, we used the resistivities $\rho_{SRO}$ of SrRu$_{1-v}$O$_3$ and $\rho_{LSMO}$ of LSMO obtained by electrical transport measurements for our samples [Fig. 3(a)]; $\rho_{SRO}$ was estimated from the experimental sheet resistance of the LSMO/SrRu$_{1-v}$O$_3$ sample and $\rho_{LSMO}$ (=3.0 mΩ cm), which was obtained by a standard four-probe method for an LSMO/STO sample grown with the same condition. As for the values of $\lambda$, we used the values obtained for SrRu$_{1-v}$O$_3$ in Ref. 52. The spin-to-charge conversion efficiency (spin Hall angle) $\theta_{\text{SHE}}$ and spin Hall conductivity $\sigma_{\text{SHE}}$ (= $\theta_{\text{SHE}}\sigma_{\text{SRO}}$) were estimated to be 0.039 and 2.1×10$^4$ ($\hbar/e$) S/m for SrRuO$_3$ and 0.0052 and 1.6×10$^3$ ($\hbar/e$) S/m for SrRu$_{0.7}$O$_3$ at 300 K, indicating that the stoichiometry condition is essential for obtaining high $\theta_{\text{SHE}}$ and $\sigma_{\text{SHE}}$.

Here, we discuss the origin of the significant decrease in $\theta_{\text{SHE}}$ for non-stoichiometric SrRu$_{0.7}$O$_3$ compared with the stoichiometric counterpart. The possible origins of the inverse spin Hall effect are side jump [10,11], skew scattering [12,13], and Karplus-Luttinger mechanism [14]. Among them, for SrRuO$_3$, the side jump and Karplus-Luttinger mechanism have been considered to mainly contribute to the spin (or anomalous) Hall effect [25,53]. Electron scattering is the basic origin of both effects. The side jump mechanism depends on spin-orbit interactions, while the Karplus-Luttinger mechanism originates from the spin-Berry curvature of the electronic band structure. Ru defects greatly influence the spin-charge interconversion in both cases. In the side jump mechanism, scattering is caused by Ru atoms with strong spin-orbit interaction. Thus, the absence of Ru atoms causes a decrease in the spin-to-charge conversion efficiency. In the Karplus-Luttinger mechanism, Ru defects affect the electronic band structure, causing a significant change in the spin-to-charge conversion efficiency. Which mechanism is dominant can generally be determined from the conductivity (carrier lifetime). In the ultraclean ragime, where carriers have a long lifetime, skew scattering is dominant. With decreasing the carrier lifetime, materials gradually transition into the clean ragime where the Karplus-Luttinger mechanism is dominant. As the carrier lifetime further decreases, materials enter a bad metal ragime, where the Karplus-Luttinger mechanism is still dominant until entering the hopping conduction ragime. SrRuO$_3$ is classified as a bad metal at room temperature (300 K) in terms of its conductivity. A previous report compared the temperature dependencies between the conductivity and the spin Hall conductivity in SrRuO$_3$ and indicated that the Karplus-Luttinger mechanism is dominant at 300 K [22]. Thus, our experimental spin-to-charge conversions are most likely attributed to the Karplus-Luttinger mechanism.

To investigate the contribution of the Karplus-Luttinger mechanism to the spin-to-charge conversion, we theoretically calculated the electronic band structures of SrRuO$_3$ and SrRu$_{0.75}$O$_3$ [Figs. 6(a) and 6(b)]. Here, we simplified the calculation by adopting SrRu$_{0.75}$O$_3$ instead of SrRu$_{0.7}$O$_3$ so that we can perform the calculation with only one unit cell of orthorhombic SrRuO$_3$ with a Ru atom removed. The spin Berry curvature $\Omega_{zx}^{y}(\boldsymbol{k})$ and spin Hall conductivity $\sigma_{zx}^{\text{SHE},y}$ of SrRu$_{1-v}$O$_3$ are

expressed as below [54]:

$$\Omega_{zx}^{n,y}(\boldsymbol{k}) = -\sum_{m\neq n} \frac{2\text{Im}[\langle n\boldsymbol{k}|j_z^{s,y}|m\boldsymbol{k}\rangle\langle m\boldsymbol{k}|v_x|n\boldsymbol{k}\rangle]}{(\varepsilon_{n\boldsymbol{k}} - \varepsilon_{m\boldsymbol{k}})^2}, \quad (5)$$

$$\Omega_{zx}^{y}(\boldsymbol{k}) = \sum_n f_{n\boldsymbol{k}} \Omega_{zx}^{n,y}(\boldsymbol{k}), \quad (6)$$

$$\sigma_{zx}^{\text{SHE},y} = e\hbar \int_{BZ} \frac{d\boldsymbol{k}}{(2\pi)^3} \Omega_{zx}^{y}(\boldsymbol{k}), \quad (7)$$

where *n* and *m* are band indices, $f_{n\boldsymbol{k}}$, $\Omega_{zx}^{n,y}(\boldsymbol{k})$, and $\varepsilon_{n\boldsymbol{k}}$ are the Fermi-Dirac distribution function, the spin Berry curvature for the *n*th band, and the *n*th eigenenergy at a wave vector $\boldsymbol{k}$, respectively. $j_z^{s,y}$ and $v_x$ are the spin-current and velocity operators, respectively. $\sigma_{zx}^{\text{SHE},y}$ is the intrinsic spin Hall conductivity when an injected spin current, which flows along the [001] direction of SrTiO$_3$ with a spin polarization along the *y* direction (//[110] of SrTiO$_3$), is converted into a charge current along the *x* direction (//[1$\bar{1}$0] of SrTiO$_3$). As shown in Fig. 6(c), when the Fermi level $E_F$ is set at 0 eV, the calculated $\Omega_{zx}^{y}(\boldsymbol{k})$ for SrRuO$_3$ has a sharp peak at $\boldsymbol{k}$ where a band crossing appears near $E_F$ in the Y−Γ path [see the black square in Fig. 6(a)]. When $E_F$ = 0, SrRuO$_3$ has larger $\Omega_{zx}^{y}(\boldsymbol{k})$ than SrRu$_{0.75}$O$_3$ in the Y−Γ path. As shown in Fig. 6(d), SrRuO$_3$ has higher $\sigma_{zx}^{\text{SHE},y}$ than SrRu$_{0.75}$O$_3$ when $E_F \approx 0$.

The experimental $\sigma_{\text{SHE}}$ and theoretical $\sigma_{zx}^{\text{SHE},y}$ at $E_F$ = 0 are summarized in Table 1. The experimental result of the decrease in the spin Hall conductivity for non-stoichiometric SrRu$_{0.7}$O$_3$ is well reproduced by our theoretical calculation. The slight difference between the theoretical and experimental results is probably due to a deviation in $E_F$ from 0 eV or the Ru composition difference between the calculations and the actual samples. We also note that the side jump scattering might also contribute slightly. Therefore, our results strongly suggest that the difference in experimental $\theta_{\text{SHE}}$ and $\sigma_{\text{SHE}}$ between SrRuO$_3$ and SrRu$_{0.7}$O$_3$ mainly originates from the spin Berry curvature. Our results mean that materials with large spin-orbit interactions and band crossings can exhibit large spin Berry curvature, leading to highly efficient spin-charge interconversion.

### IV. SUMMARY

We have investigated how the deviation in the Ru composition from the stoichiometric condition affects the spin-to-charge conversion in SrRuO$_3$ by carrying out spin-pumping measurements for SrRuO$_3$ and SrRu$_{0.7}$O$_3$ films. The obtained voltage due to the spin-to-change conversion in SrRuO$_3$ was higher than in SrRu$_{0.7}$O$_3$. Our analysis based on this experimental result indicates that $SrRuO_3$ has a higher spin-Hall angle (0.039) than $SrRu_{0.7}O_3$ (0.0052). We calculated the band structures of $SrRuO_3$ and $SrRu_{0.75}O_3$ using first-principles calculations to theoretically estimate the spin-Hall conductivity originating from the spin Berry curvature, revealing the decrease in the spin-Hall conductivity in non-stoichiometric $SrRu_{0.7}O_3$ in comparison with $SrRuO_3$. This theoretical result agrees well with the experimental results. Our results suggest that the composition deviation in materials largely changes the spin Berry curvature and the spin-charge interconversion efficiency. These insights will help us with material designs for realizing efficient spin-charge interconversion.

## Data availability

The data that support the findings of this study are available from the corresponding author upon reasonable request.

## Acknowledgments

This work was partly supported by Grants-in-Aid for Scientific Research (No. 20H05650, 21J21102, 22H04948, 23H03802, 23H03805), CREST (No. JPMJCR1777), and ERATO (No. JPMJER2202) of the Japan Science and Technology Agency, the Spintronics Research Network of Japan (Spin-RNJ), and the ANRI fellowship. Part of this work was conducted at the Advanced Characterization Nanotechnology Platform of the University of Tokyo, and it was supported by the "Nanotechnology Platform" of the Ministry of Education, Culture, Sports, Science and Technology (MEXT), Japan, and the Cryogenic Research Center of the University of Tokyo. S.K.T. acknowledges support from the Japan Society for the Promotion of Science (JSPS) Fellowships for Young Scientists.

## Author Contributions

Experimental design, sample growth, measurement, and data analysis: S.K.T.; $SrRu_{1-v}O_3$ growth: Y.K.W. and Y.K.; theoretical calculation: S.K.T. and H.S.; writing and project planning: S.K.T. and S.O.; S.K.T., Y.K.W., H.S., Y.T., H.Y., Y.K., M.T., and S.O. discussed the results and the manuscript extensively.

## Competing interests
The authors declare no competing interests.

effect in metallic ferromagnets, Phys. Rev. B **88**, 125110 (2013).

17. K. Takiguchi, Y. K. Wakabayashi, H. Irie, Y. Krockenberger, T. Otsuka, H. Sawada, S. A. Nikolaev, H. Das, M. Tanaka, Y. Taniyasu *et al.*, Quantum transport evidence of Weyl fermions in an epitaxial ferromagnetic oxide, Nat. Commun. **11**, 4969 (2020).

18. S. Kaneta-Takada, Y. K. Wakabayashi, Y. Krockenberger, S. Ohya, M. Tanaka, Y. Taniyasu, H. Yamamoto, High-mobility two-dimensional carriers from surface Fermi arcs in magnetic Weyl semimetal films, npj Quantum Mater. **7**, 102 (2022).

19. U. Kar, A. K. Singh, Y. -T. Hsu, C. -Y. Lin, B. Das, C. -T. Cheng, M. Berben, S. Yang, C. -Y. Lin, C. -H. Hsu *et al.*, The thickness dependence of quantum oscillations in ferromagnetic Weyl metal $SrRuO_3$, npj Quantum Mater. **8**, 8 (2023).

20. Z. Fang, N. Nagaosa, K. S. Takahashi, A. Asamitsu, R. Mathieu, T. Ogasawara, H. Yamada, M. Kawasaki, Y. Tokura, and K. Terakura, The Anomalous Hall Effect and Magnetic Monopoles in Momentum Space, Science **302**, 92−95 (2003).

21. R. Mathieu, A. Asamitsu, H. Yamada, K. S. Takahashi, M. Kawasaki, Z. Fang, N. Nagaosa, and Y. Tokura, Scaling of the Anomalous Hall Effect in $Sr_{1-x}Ca_xRuO_3$, Phys. Rev. Lett. **93**, 016602 (2004).

22. Y. Ou, Z. Wang, C. S. Chang, H. P. Nair, H. Paik, N. Reynolds, D. C. Ralph, D. A. Muller, D. G. Schlom, and R. A. Buhrman, Exceptionally High, Strongly Temperature Dependent, Spin Hall Conductivity of $SrRuO_3$, Nano. Lett. **19**, 3663−3670 (2019).

23. S. Kaneta-Takada, Y. K. Wakabayashi, Y. Krockenberger, S. Ohya, M. Tanaka, Y. Taniyasu, and H. Yamamoto, Thickness-dependent quantum transport of Weyl fermions in ultra-high-quality $SrRuO_3$ films, Appl. Phys. Lett. **118**, 092408 (2021).

24. Y. K. Wakabayashi, S. Kaneta-Takada, Y. Krockenberger, S. Ohya, M. Tanaka, Y. Taniyasu, and H. Yamamoto, Structural and transport properties of highly Ru-deficient $SrRu_{0.7}O_3$ thin films prepared by molecular beam epitaxy: Comparison with stoichiometric $SrRuO_3$, AIP Advances **11**, 035226 (2021).

25. S. Kaneta-Takada, Y. K. Wakabayashi, Y. Krockenberger, H. Irie, S. Ohya, M. Tanaka, Y. Taniyasu, and H. Yamamoto, Scattering-dependent transport of $SrRuO_3$ films: From Weyl fermion transport to hump-like Hall effect anomaly, Phys. Rev. Matter. **7**, 054406 (2023).

26. Y. K. Wakabayashi, T. Otsuka, Y. Krockenberger, H. Sawada, Y. Taniyasu, and H. Yamamoto, Machine-learning-assisted thin-film growth: Bayesian optimization in molecular beam epitaxy of $SrRuO_3$ thin films, APL Mater. **7**, 101114 (2019).

27. Y. K. Wakabayashi, T. Otsuka, Y. Krockenberger, H. Sawada, Y. Taniyasu, and H. Yamamoto, Bayesian optimization with experimental failure for high-throughput

**Figures and Captions**

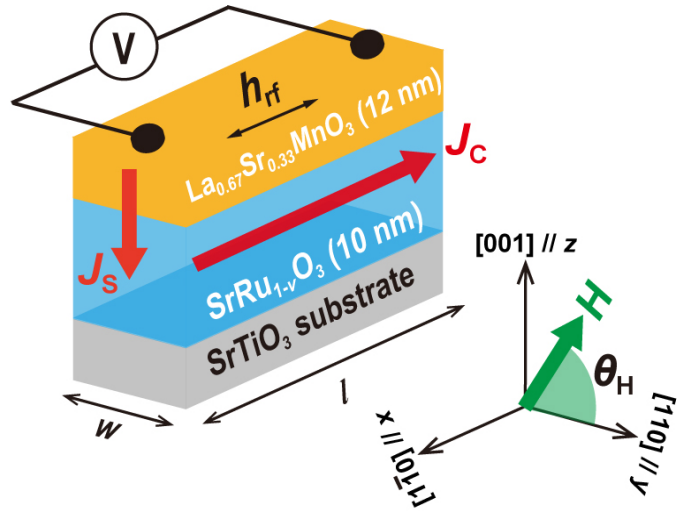

FIG. 1. Schematic structure of the (La$_{0.67}$,Sr$_{0.33}$)MnO$_3$/SrRu$_{1-v}$O$_3$ ($v$ = 0 and 0.3)/SrTiO$_3$ samples used for spin pumping measurements.

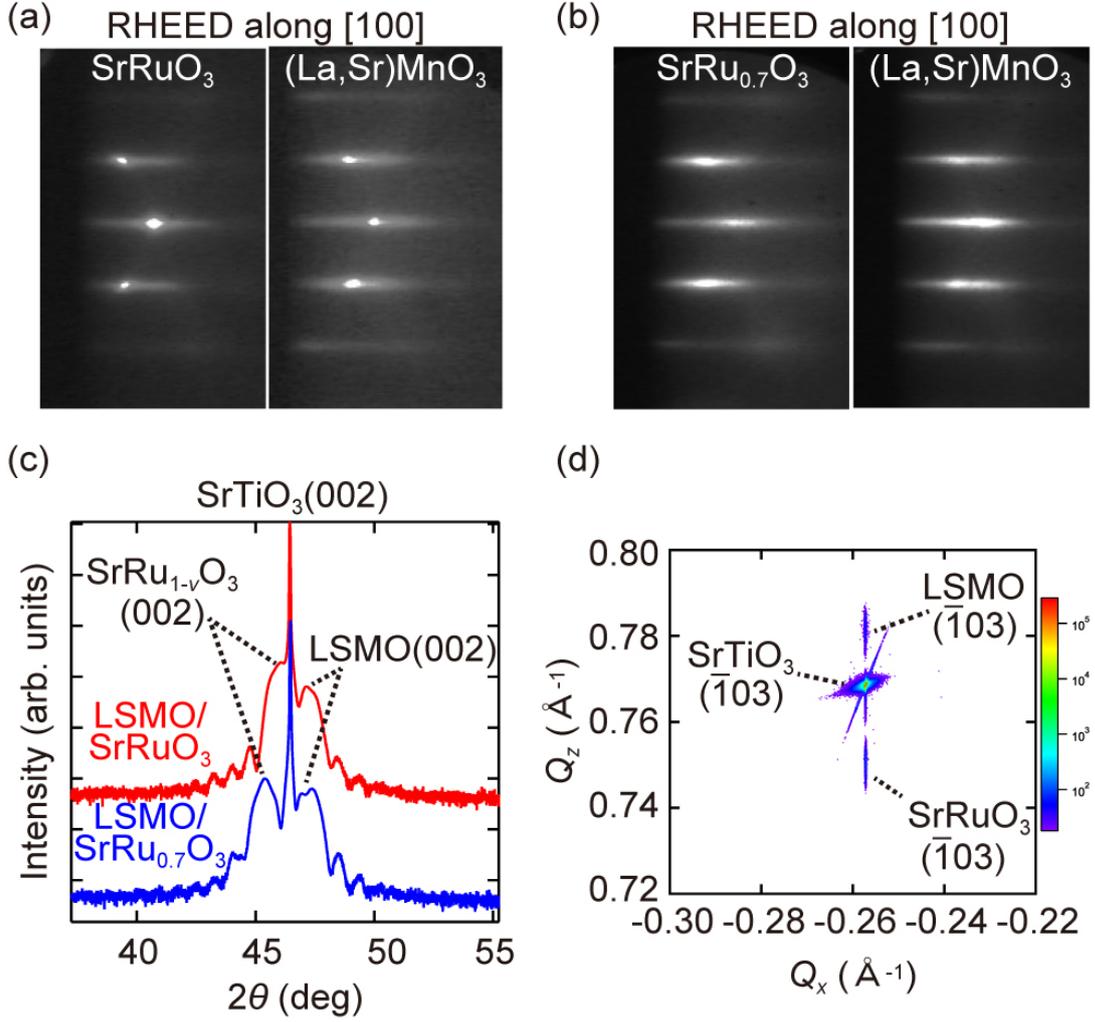

FIG. 2. (a) and (b) *In-situ* RHEED patterns obtained for the (a) LSMO/SrRuO$_3$ and (b) LSMO/SrRu$_{0.7}$O$_3$ samples with the incident electron beam azimuth along the [100] direction of the substrate. (c) The logarithm intensity plot of $\theta$-$2\theta$ XRD patterns for the LSMO/SrRuO$_3$ and LSMO/SrRu$_{0.7}$O$_3$ samples. Offsets are multiplied for easy comparison. (d) Reciprocal space mapping of LSMO/SrRuO$_3$ around ($\bar{1}$03) of the SrTiO$_3$ substrate. $Q_z$ and $Q_x$ represent the components of the reciprocal lattice vector in the direction perpendicular to the film plane and the in-plane [100] direction, respectively.

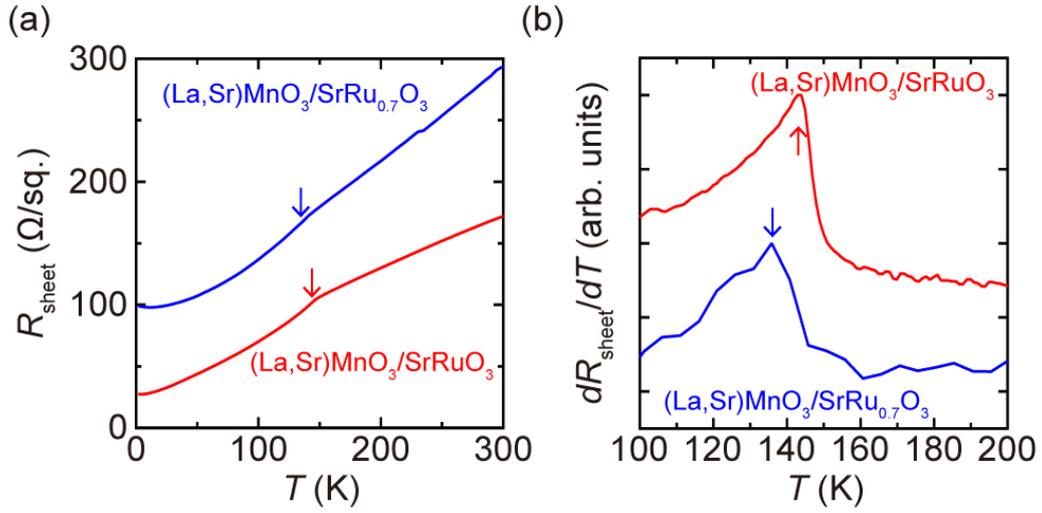

FIG. 3. (a) $T$ dependence of the sheet resistance $R_{sheet}$ of the LSMO/SrRuO$_3$ and LSMO/SrRu$_{0.7}$O$_3$ samples. (b) $T$ dependence of the differential sheet resistance $dR_{sheet}/dT$ for the LSMO/SrRuO$_3$ and LSMO/SrRu$_{0.7}$O$_3$ samples. In (a) and (b), the arrows indicate the positions of the peaks or kinks. In (b), $dR_{sheet}/dT$ is offset for easy viewing.

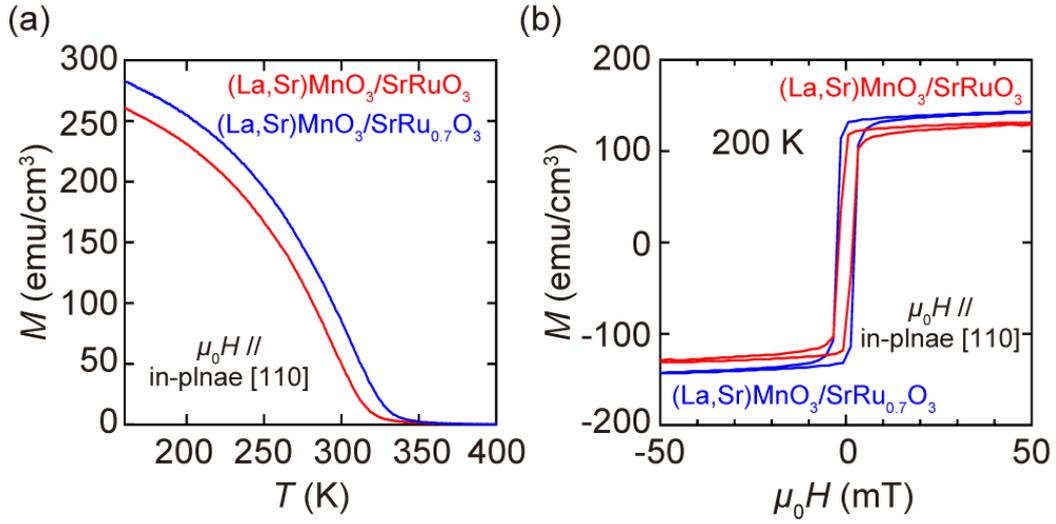

FIG. 4. (a) $T$ dependence of $M$ of LSMO/SrRuO$_3$ and LSMO/SrRu$_{0.7}$O$_3$ with a magnetic field $\mu_0H$ of 100 mT applied along the in-plane [110] axis. (b) $M$ as a function of $\mu_0H$ for LSMO/SrRuO$_3$ and LSMO/SrRu$_{0.7}$O$_3$ measured at 200 K with a magnetic field applied along the in-plane [110] direction of the SrTiO$_3$ substrate. The measurement temperature of 200 K is higher than $T_C$ of SrRu$_{1-v}$O$_3$ ($v$ = 0, 0.3).

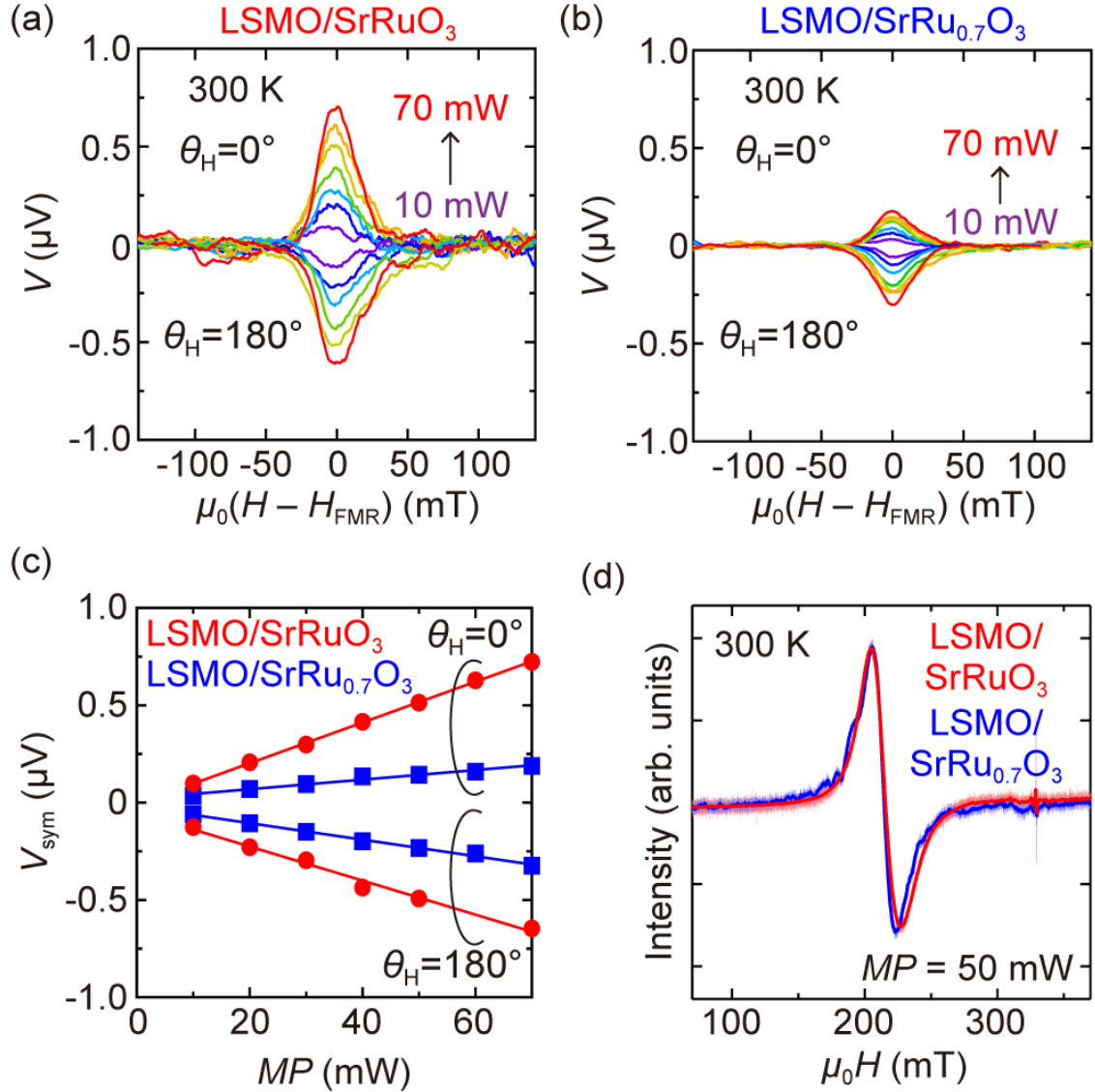

FIG. 5. (a) and (b) $\mu_0 H$ dependencies of $V$ at 300 K measured for (a) LSMO/SrRuO$_3$ and (b) LSMO/SrRu$_{0.7}$O$_3$ with various $MP$ values ranging from 10 to 70 mW. In the electron spin resonance system, $h_{rf}$ was applied along the [1$\bar{1}$0] direction of the SrTiO$_3$ substrate. (c) Microwave power ($MP$) dependences of the symmetric component $V_{sym}$ in LSMO/SrRuO$_3$ and LSMO/SrRu$_{0.7}$O$_3$ for $\theta_H = 0°$ and 180°. (d) Magnetic field $\mu_0 H$ (// [110] of the SrTiO$_3$ substrate) dependence of the microwave absorption derivative for the LSMO/SrRuO$_3$ and LSMO/SrRu$_{0.7}$O$_3$ samples at 300 K with $MP = 50$ mW. For easy viewing, we used data after smoothing. The dark and light red/blue lines are smoothed and raw data.

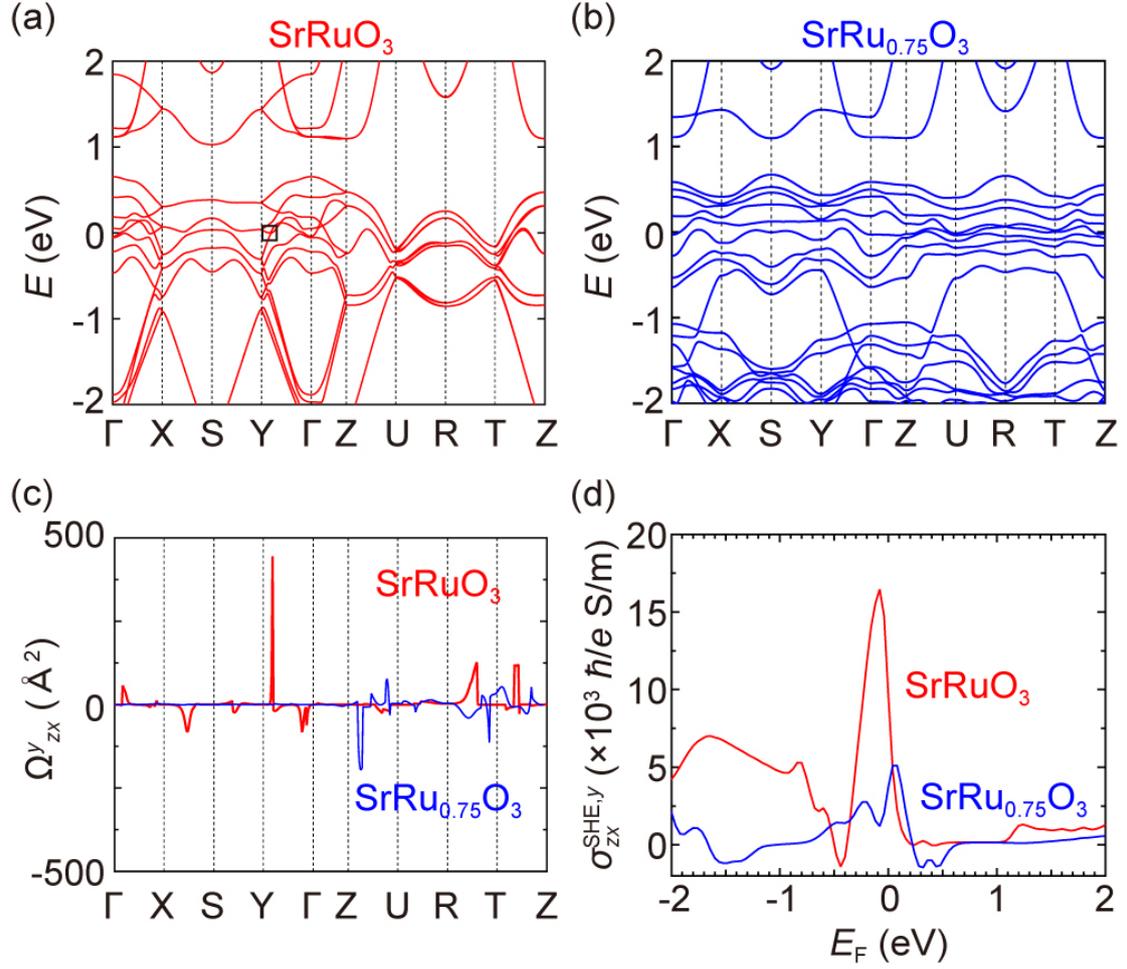

FIG. 6. (a) and (b) Calculated band structures with spin-orbit coupling for (a) SrRuO$_3$ and (b) SrRu$_{0.75}$O$_3$. The black square in (a) indicates a band crossing point, where the spin-charge interconversion occurs in SrRuO$_3$. (c) Comparison of the $\boldsymbol{k}$-resolved spin Berry curvatures $\Omega_{zx}^{y}(\boldsymbol{k})$ of SrRuO$_3$ and SrRu$_{0.75}$O$_3$. (d) $E_F$ dependence of the calculated spin Hall conductivity $\sigma_{zx}^{\mathrm{SHE},y}$ of SrRuO$_3$ and SrRu$_{0.75}$O$_3$.

**Table and Caption**

TABLE 1 Comparison between the experimental $\sigma_{\text{SHE}}$ and theoretical $\sigma_{zx}^{\text{SHE},y}$ for SrRuO$_3$ and SrRu$_{0.7}$O$_3$ at 300 K.

|  | SrRuO$_3$ | SrRu$_{0.7}$O$_3$ (SrRu$_{0.75}$O$_3$) |
| --- | --- | --- |
| $\sigma_{\text{SHE}}$ ($\hbar/e$) S/m | 2.1×10$^4$ | 1.6×10$^3$ |
| $\sigma_{zx}^{\text{SHE},y}$ ($\hbar/e$) S/m | 9.6×10$^3$ | 3.7×10$^3$ |